\documentstyle[preprint,aps,floats,epsfig]{revtex}
\begin{document}

\begin{center}
{\Large {\bf Wave packet propagation study of the charge transfer\\[0pt]
interaction in the F${}^-$ -- Cu(111) and -- Ag(111) systems}}

\vskip 1cm {\bf A.G. Borisov, J.P. Gauyacq}

\vskip 0.3cm {\it Laboratoire des Collisions Atomiques et Mol\'eculaires,\\[%
0pt]
Unit\'e mixte de recherche CNRS-Universit\'e Paris-Sud UMR 8625,\\[0pt]
B\^atiment 351, Universit\'e Paris-Sud, 91405 Orsay CEDEX, France}

\vskip 0.5cm

{\bf S.V. Shabanov}

\vskip0.3cm {\it Department of Mathematics, University of Florida, Little
Hall 358,\\[0pt]
Gainesville, FL 32611, USA}

\bigskip \bigskip \bigskip \bigskip \bigskip

Abstract
\end{center}

The electron transfer between an $F^{-}$ ion and $Cu(111)$ and $Ag(111)$
surfaces is studied by the wave packet propagation method in order to
determine specifics of the charge transfer interaction between the negative
ion and the metal surface due to the projected band gap. A new modeling of
the $F^{-}$ ion is developed that allows one to take into account the six
quasi-equivalent electrons of $F^{-}$ which are {\it a priori} active in the
charge transfer process. The new model invokes methods of constrained
quantum dynamics. The six-electron problem is transformed to two
one-electron problems linked via a constraint. The projection method is used
to develop a wave packet propagation subject to the modeling constraint. The
characteristics (energy and width) of the ion $F^{-}$ ion level interacting
with the two surfaces are determined and discussed in connection with the
surface projected band gap.

\section{ Introduction}

When an atom or a molecule is close to a surface of a solid, its electrons
interact with those of the solid, leading to the possibility of an electron
transfer between the atom (molecule) and the solid. This charge transfer
process plays a very important role in a variety of different situations. In
particular, it often occurs as an intermediate step in reactions at surfaces
(desorption, fragmentation of adsorbates, chemical reactions, quenching of
metastable species, etc.) \cite{1,2,3,4}. The one-electron transfer between
energetically degenerate electronic levels of the atom and the solid is
called the Resonant Charge Transfer (RCT) process. It is usually considered
as the most efficient one among various possible charge transfer
interactions. Since a few years the development of accurate theoretical
approaches to the RCT in the case of free electron metal surfaces \cite
{5,6,7,8,9,10,11} has led to a successful description of a one-electron
transfer in the interaction of ions (atoms) with such surfaces \cite
{12,13,14}. All these approaches are based on the description of single
electron being transferred between the atom and the surface. As an example,
one can mention the neutralisation of an alkali positive ion by RCT, even if
the possibility of capturing the electron in two different spin states
somewhat alters the one-electron picture \cite{15,16}. However, many atoms
or molecules contain more than one electron which can possibly participate
in the charge transfer process, especially, when the active electrons occupy
the same energy level. The latter has to be taken into account in any
quantitative approach to the RCT \cite{12,13,14,15,16,17}.

For instance, a free fluorine negative ion can be described as a closed $%
2p^{6}$ outer electronic shell with six equivalent electrons, and all of
them can participate in the RCT when the ion is close to a metal surface.
The effects of each electron cannot be simply added up to get a total
effect. Indeed, consider the loss of an electron by the ion. Any of the six
electrons can be the one that is detached. However, the detachment of one
electron {\it a priori} precludes the detachment of another one. Clearly,
the loss of a second electron would be a completely different process
because it corresponds to a formation of a positive fluorine ion and, hence,
has a different energetics. Such correlations between the outer-shell
electrons of the $F^{-}$ ion must be accounted for in any description of the
RCT. In this work we show how such correlations can be modeled by two
one-electron systems linked via a {\it constraint}.

In classical dynamics, constraints appear as some algebraic relations
between generalized coordinates of a system and their time derivatives
(velocities), which are to hold for any moment of time. They are widely used
to model, e.g., effects of an environment on a system in question, or to
develop theories with local symmetries (gauge theories) such as
electrodynamics, general relativity and Yang-Mills theories. Classical
constrained dynamics has been well studied in classical works by d'Alambert,
Lagrange, H\"{o}lder and Gauss (see, e.g., Ref. \cite{18} and references
therein). In quantum mechanics constraints have been analyzed by Dirac \cite
{19}, mainly because of the need of quantizing fundamental physical theories
with local symmetries. The question addressed by Dirac was the following: 
{\it Given a classical system with constraints, construct a quantum theory
which satisfies the correspondence principle}. A time evolution of a quantum
system is described by the evolution operator. Its kernel in the coordinate
representation is called a quantum mechanical propagator. For constrained
systems it is usually obtained by a reduction of the Feynman phase space
path integral onto the physical phase space, as proposed by Faddeev \cite{20}%
. There are many subtleties associated with quantization of constrained
systems (see for a review, e.g., \cite{21}).

Here we propose a quantum constrained system which can be used to model the
charge transfer interaction between an $F^{-}$ ion and a metal surface. The
basic idea is to transform the original six-electron problem to two
one-electron problems each of which describes a possible decay channel of
the ion. As a consequence of the quantum equivalence of the outer-shell
electrons, the effective one-electron systems turn out to be linked {\em only%
} by a constraint. In other words, the entire interaction between the two
systems occurs through a constraint rather than via a local potential. One
can also say that such a kinematic coupling of the decay channels is induced
by the symmerty of the original six-electron problem.

The constraint has a remarkable feature: It does not have a classical limit,
meaning that its effects disappear in the formal classical limit $\hbar
\rightarrow 0$. That is, the dynamical effect modeled by such a constraint
is essentially quantum and cannot be modeled by any classical potential
force. In this regard, the canonical quantization scheme for constrained
systems due to Dirac is not applicable here. We develop below a novel
computational scheme for a quantum evolution subject to such purely quantum
constraints. The scheme is based on the projection formalism introduced
earlier by one of us \cite{22} within the framework of gauge theories. Our
approach also comprises a novel method to account for the intra-atomic
correlations within a one-electron description of the charge transfer
interaction between an $F^{-}$ ion and the $Cu(111)$ and $Ag(111)$ surfaces.
We demonstrate that the approach turns out to be very efficient in the wave
packet propagation studies.

The choice of the $F^{-}/Cu(111)$ and $F^{-}/Ag(111)$ systems is motivated
by several reasons. The interaction of halogen negative ions with a
free-electron metal surface has already been studied theoretically within
the Coupled Angular Mode approach which lead to a quite satisfactory
description of the halogen negative ion formation in scattering halogen
atoms by various metal surfaces \cite{12,13}. Recently, on the example of
the $Cu(111)$ surface it has been shown that the projected band gap of the $%
(111)$ surfaces of noble metals strongly affects the RCT \cite{23,24}. The
point is that in a certain energy range electrons cannot propagate along the
normal to the surface (L-band gap in the $<111>$ direction \cite{25}). On
the other hand, the RCT process corresponds to an electron tunneling between
the atom and the surface which is favored along the surface normal. In
addition, it was shown that the 2D electronic continuum of the surface state
plays a significant role, often dominating the RCT process. The band gap has
several consequences. First, there exist very long lived states in the
alkali--$Cu(111)$ systems \cite{24,26,27}. Second, there is a parallel
velocity dependence of the probability of an electron capture by the atom
from $Cu(111)$ surfaces in grazing scattering experiments that is
characteristic of a 2D electronic continuum \cite{28}. Third, there exits an
avoided crossing between the energy level of the projectile and the bottom
of the surface state continuum \cite{23}. These effects have been found to
depend strongly on the interaction time \cite{23,29,30}: They only appear
for long interaction times, i.e., for slow collisions.

The free $F^{-}$ ion energy level is slightly above the bottom of the
surface state continuum of the $Cu(111)$ and $Ag(111)$ surfaces, and
therefore, because of the image charge attraction, it could cross the bottom
of the 2D continuum for a finite ion-surface distance. In the present work,
we investigate the effects of the projected band gap and of the surface
state continuum on the $F^{-}$-metal charge transfer. In particular, we
study the behavior of the system when the ion energy level is very close to
the bottom of the 2D surface state continuum. A Wave Packet Propagation
(WPP) approach to the RCT \cite{8,23} provides a quantitative description of
dynamic and static aspects of the $F^{-}$-surface charge transfer. The
latter allows one to analyze the dependence of the band structure effect
upon an interaction time in the RCT.

\section{Method}

\subsection{A negative ion $F^-$}

The free negative ion $F^{-}$ is usually described in the Hartree--Fock
approximation as a closed shell ion with the electronic configuration $%
1s^{2}2s^{2}2p^{6}$. Its binding energy is 3.4 eV. In this approach the
outer shell electrons are regarded as equivalent. Their wave function is
given by the corresponding Slater determinant 
\begin{equation}
\Phi =\frac{1}{\sqrt{6!}}\left| 2p_{0}^{\alpha }2p_{0}^{\beta
}2p_{1}^{\alpha }2p_{1}^{\beta }2p_{-1}^{\alpha }2p_{-1}^{\beta }\right| \ ,
\label{1}
\end{equation}
where $2p$ symbolizes a $2p$ orbital wave function, the superscripts $\alpha 
$ and $\beta $ stand for the two possible spin directions, and the subscript
indicates the magnetic quantum number $m$ corresponding to the projection of
the $2p$ angular momentum on the quantization axis. The Slater determinant (%
\ref{1}) can be expanded into a sum of products of wave functions of an
ionic core and an outer electron: 
\begin{equation}
\Phi =\frac{1}{\sqrt{6}}\sum_{j=1}^{6}\left| \,F_{j}\,\right| A_{j}\phi
_{j}=\sum_{j=1}^{6}\left| \,G_{j}\,\right| \phi _{j}\ ,  \label{2}
\end{equation}
where the five-electron determinant $|\ F_{j}|$ describes a state of the
fluorine atom $F({}^{2}P)$. The product $A_{j}\phi _{j}$ corresponds to the $%
2p_{m}^{\alpha ,\beta }$ orbital, which has been factorised into a spin
factor $A_{j}$ and a spatial wave function $\phi _{j}=\phi _{j}(\vec{r}\,)$.
The wave function (\ref{2}) can also be used for an open shell description
of the negative ion of the type $2p^{5}2p^{\prime }$. In this case, the
spatial wave function $\phi _{j}$ singled out in (\ref{2}) corresponds to
the outer $2p^{\prime }$ orbital of the negative ion, whereas the core wave
function $|G_{j}|$ is formed by the inner $2p$ orbitals. When analyzing the
electron detachment process in the open shell description, the $2p^{\prime }$
orbital is regarded as an active one.

The representation (\ref{2}) is particularly well suited for an analysis of
the electron capture/loss process between a fluorine ion $F^{-}$ and a metal
surface since it gives an expansion of the ion wave function over possible
detachment channels. So, we retain the representation (\ref{2}) to describe
a loss or capture of an electron by the fluorine core 
\begin{equation}
\Psi =\sum_{j}\left| \,G_{j}\,\right| \psi _{j}\ .  \label{3}
\end{equation}
Here $\psi _{j}=\psi _{j}(\vec{r}\,)$ is a wave function of an electron
(captured or lost) in the $j$-channel. Neglecting the spin-orbit
interactions and assuming the ion-surface interaction to be invariant under
translations of the ion parallel to the surface, the system becomes
invariant under the spin flip and rotations about the $z$-axis which is set
to be perpendicular to the metal surface and passing through the ion center.
Next, the $z$-axis is chosen as the quantization axis. Therefore the states
with the $z$-component of the electron angular momentum $\pm 1$ are
degenerate. As a consequence, only two electron wave functions are distinct
in the representation (\ref{3}). They correspond to the states with $m=0$
and $|m|=1$. In what follows we assume that the charge transfer does not
affect the neutral core wave function $|G_{j}|$. Only the outer electron is
subject to the RCT dynamics.

Any state of the system can be represented as a two dimensional isovector $%
\left| \Psi \right\rangle $ whose components are one-electron states
corresponding to $m=0$ and $|m|=1$ (upper and lower elements of the
isovector, respectively). In cylindrical coordinates $(z,\rho ,\varphi )$,
it can be written as 
\begin{equation}
\left\langle \vec{r}\text{ }|\Psi \right\rangle =\left( 
\begin{array}{c}
\psi _{0}(z,\rho ) \\ 
\psi _{1}(z,\rho )e^{i\varphi }
\end{array}
\right) \ .  \label{4}
\end{equation}
Note that, due to the symmetry of the problem, the $\varphi $ dependence can
be explicitly given. Hence, one can limit oneself to studying the $z$- and $%
\rho $-dependence of the electron wave packet. In our representation a free
ion wave function is given by 
\begin{equation}
\left\langle \vec{r}\text{ }|\Psi _{F^{-}}\right\rangle =\left( 
\begin{array}{c}
\frac{1}{\sqrt{3}}p(r)\ Y_{10}(\theta ,\varphi ) \\ 
\sqrt{\frac{2}{3}}p(r)\ Y_{11}(\theta ,\varphi )
\end{array}
\right)  \label{5}
\end{equation}
where $p(r)$ is the radial part of the free-ion orbital and $Y_{lm}(\theta
,\varphi )$ are the spherical harmonics. The different normalization factors
of the isovector components are due to the fact that the free ion wave
function (\ref{2}) contains twice as many states with $|m|=1$ as with $m=0$.
The functions $p(r)Y_{lm}(\theta ,\varphi )$ have a unit norm so that $%
\left\langle \Psi _{F^{-}}|\Psi _{F^{-}}\right\rangle =1$.

Yet another remark is that the approach outlined above assumes that only an
outer electron can be detached (the detachment occurs through the evolution
of the wave functions $\psi_0$ and $\psi_1$), i.e., it assumes an open shell
description ($2p^52p^\prime$) of the ion $F^-$. A closed shell description ($%
2p^6$) would correspond to an increase of the charge transfer interaction by
the factor $\sqrt{6}$, and, hence, to an increase of the width by the factor
6 (see a discussion in \cite{16}). This appears to be better adapted for the
halogen negative ion case.

With all the above settings, a modeling of the charge transfer in the $F^{-}$%
-metal system implies finding a one-electron Hamiltonian that governs a time
evolution in the Hilbert space spanned by vectors (\ref{4}). We take it in
the following form 
\begin{equation}
{\bf H}={\bf T}+{\bf V}_{{\rm at}}+{\bf V}_{{\rm S}}={\bf H}_{{\rm at}}+{\bf %
V}_{{\rm S}}\ .  \label{6}
\end{equation}
The operators ${\bf H},\ {\bf T},\ {\bf V}_{{\rm at}},\ {\bf V}_{{\rm S}}$
and ${\bf H}_{{\rm at}}$ are diagonal 2$\times $2 matrices, in fact, we
choose them to be proportional to a unit 2$\times $2 matrix, with the
diagonal elements denoted, respectively, $H,\ T,\ V_{{\rm at}},\ V_{{\rm S}}$
and $H_{{\rm at}}$. Here, $T$ is the electron kinetic energy operator, $V_{%
{\rm at}}$ the potential of the interaction between an electron and the
neutral core, $V_{{\rm S}}$ the potential of the electron-surface
interaction, $H_{{\rm at}}=T+V_{{\rm at}}$, and $H=H_{{\rm at}}+V_{{\rm S}}$%
. The wave functions $p(r)Y_{lm}(\theta ,\varphi )$ are eigenfunctions of
the one-electron Hamiltonian $H_{{\rm at}}$.

The time evolution of the wave function (\ref{4}) generated by the
Hamiltonian (\ref{6}) is nothing but the evolution of two independent
one-electron wave packets. On the other hand, the two components of the
isovector (\ref{4}) cannot evolve independently. Indeed, in the free ion
case, the two components cannot be arbitrarily chosen in order for the state
(\ref{4}) to describe a physical free ion. Actually, the radial components
of the column elements appear to be proportional to each other with a
specific factor (cf. (\ref{5})). This relation between the two components
comes from the expansion of the Slater determinant (\ref{1}) which possesses
a high symmetry being the symmetry of a quantum system of identical
particles occupying the same energy level. The very same symmetry must be
preserved in the expansion (\ref{2}) and, hence, upon a reduction of the
six-electron description to our one-electron formalism. In other words,
there must be a correlation between evolving components of the isovector
corresponding to the electron states with $m=0$ and $|m|=1$ thanks to the
symmetry of the six-electron problem. Therefore physically admissible states
in the Hilbert space spanned by isovectors (\ref{4}) must be subject to some
constraints required by the symmetry of the original six-electron problem.
This will be a key point of our new approach to the RCT dynamics.

To illustrate the necessity of constraints, consider the case when the
detachment occurs in the $m=0$ channel. Then the bound part of the $|m|=1$
channel must also disappear at the same time because there is only one ion $%
F^{-}$ which contains both $m=0$ and $|m|=1$ components. In the Effective
Range approach \cite{31} used in the Coupled Angular Mode (CAM) method \cite
{6}, this problem has been solved in the following way. One only considers
the wave function of the system given by expression (\ref{4}) outside a
spherical region of radius $r_{c}$. The $F$ core is contained in the region.
The boundary condition on the radial components $\psi _{0}$ and $\psi _{1}$
at $r=r_{c}$ couples the two channels. This approach essentially relies on
the use of spherical coordinates. Here we look for a coordinate independent
description of the channel mixing that can be {\it efficiently implemented
in numerical calculations of the time evolution of the system (wave packet
propagation)}. The origin of the kinematic coupling of the channels is now
sought in symmetries of the system. It is believed that such an approach is
rather general and could be applied to other many-electron systems where a
conventional mean field approach does not provide a good approximation.

The basic idea is that the Hilbert space spanned by isovectors (\ref{4}) is
too large and contains states which are physically not acceptable. It is
clearly seen already from the fact that the radial components of the free
ion state (\ref{5}) are not independent. We shall then constrain the state (%
\ref{4}) to allow only one bound ion $F^{-}$ of the form (\ref{5}). Any
other state with components $\psi _{0}\sim p(r)Y_{10}(\theta ,\varphi )$ and 
$\psi _{1}\sim p(r)Y_{11}(\theta ,\varphi )$ should be forbidden. This
corresponds to making the state (\ref{4}) orthogonal to the vector 
\begin{equation}
\left\langle \vec{r}\,|Q\right\rangle =\left( 
\begin{array}{c}
\sqrt{\frac{2}{3}}\,p(r)Y_{10}(\theta ,\varphi ) \\ 
-\sqrt{\frac{1}{3}}\,p(r)Y_{11}(\theta ,\varphi )
\end{array}
\right) \equiv \left( 
\begin{array}{c}
p_{0}(z,\rho ) \\ 
-p_{1}(z,\rho )e^{i\varphi }
\end{array}
\right) \ .  \label{7}
\end{equation}
Since the state (\ref{7}) cannot also occur as a virtual (or intermediate)
state of the ion in the time evolution of the system, we demand that the
time dependent wave packet (\ref{4}) must be orthogonal to the vector $%
\left| Q\right\rangle $: 
\begin{equation}
\left\langle Q|\Psi (t)\right\rangle =\left\langle p_{0}|\psi
_{0}(t)\right\rangle -\left\langle p_{1}|\psi _{1}(t)\right\rangle =0\ ,
\label{8}
\end{equation}
for any $t\geq 0$, where $\left\langle p_{0,1}|\psi _{0,1}(t)\right\rangle $
stands for a standard scalar product (written in cylindrical coordinates
because the functions $p_{0,1}$ and $\psi _{0,1}$ depend on $z$ and $\rho $
only).

{}From the physical point of view the constraint (\ref{8}) simply means that
there is an unwanted scattering mode in our effective one-electron problem.
Although the Hamiltonian may allow for such a mode, we have given physical
reasons to forbid it. The constraint (\ref{8}) implies that the time
evolution of the two components of the isovector (\ref{4}) is no longer
independent even though the Hamiltonian (\ref{6}) does not provide any
direct coupling of them. The link between the detachment channels with $m=0$
and $|m|=1$ steams directly from the free ion structure. It implicitly
assumes that the correlation between the wave functions $\psi _{0}$ and $%
\psi _{1}$ in the ion perturbed by an interaction with a metal surface is
the same as in the free ion. Thus, the electronic structure of $F^{-}$ has
been modeled by two one-electron problems linked by the constraint (\ref{8}).

\subsection{Physical Hamiltonian}

Now we face a problem of incorporating the constraint into quantum dynamics
generated by the Hamiltonian (\ref{6}). The difficulty is clear. Suppose an
initial state satisfies the constraint (\ref{8}). Applying the evolution
operator $\exp (-it{\bf H})$ to it, we immediately observe that the evolved
state fails to satisfy the constraint. The procedure we propose is based on
the projection operator formalism first introduced for gauge theories \cite
{22,32}. It has been generalized to general constrained systems \cite{33,34}
(see also the review \cite{21}). The key steps are as follows.

Consider the projection operator 
\begin{equation}
{\bf P}={\bf I}-\left| Q\right\rangle \left\langle Q\right| ,  \label{9}
\end{equation}
where ${\bf I}$ is the unit operator, ${\bf I}$ $\left| \Psi \right\rangle
=\left| \Psi \right\rangle $ for any $\left| \Psi \right\rangle $. It is
easy to convince oneself that the operator (\ref{9}) is self-adjoint, ${\bf P%
}^{\dagger }={\bf P}$, and satisfies the characteristic property of a
projection operator, ${\bf P}^{2}={\bf P}$. By construction, the state ${\bf %
P}\left| \Psi \right\rangle $ satisfies the constraint (\ref{8}) for any
state $\left| \Psi \right\rangle $. The operator (\ref{9}) projects any
state to the physical subspace defined by the condition (\ref{8}). In
particular, ${\bf P}\left| Q\right\rangle =0$. To eliminate the state $%
\left| Q\right\rangle $ as a possible intermediate state of the system in
the time evolution, the Hamiltonian is projected onto the physical subspace 
\begin{equation}
{\bf H}\ \rightarrow {\bf PHP}={\bf H}_{{\rm phys}}\ .  \label{10}
\end{equation}
The physical Hamiltonian (\ref{10}) is self-adjoint. Hence the time
evolution generated by it is unitary. The state ${\bf PHP}\left| \Psi
\right\rangle $ satisfies the constraint (\ref{8}). Clearly, the physical
Hamiltonian is nonlocal, in general, even if the original Hamiltonian has a
standard form of the sum of potential and kinetic energies. However,
classical limits of ${\bf H}$ and ${\bf H}_{{\rm phys}}$ are the same.

The evolution operator has the form 
\begin{equation}
{\bf U}(t_{1},t_{2})={\bf P}{\rm T}\exp \left( -i\int_{t_{1}}^{t_{2}}d\tau 
{\bf PHP}\right) {\bf P}={\bf P}\exp \left( -it{\bf PHP}\right) {\bf P}\ ,
\label{11}
\end{equation}
where $t=t_{2}-t_{1}$ and ${\rm T}\exp $ stands for the time ordered
exponential. The second equality holds when the Hamiltonian ${\bf H}$ does
not explicitly depend on time. The projection operators before and after the
exponential in (\ref{11}) can be omitted if the initial state satisfies the
constraint (\ref{8}).

The physical Hamiltonian provides the sought-for channel mixing. To find
terms in the new Hamiltonian which give rise to the channel mixing, we
compute the action of ${\bf PHP}$ on a generic state $\left| \Psi
\right\rangle $. A straightforward computation leads to the following result 
\begin{eqnarray}
{\bf PHP}\left| \Psi \right\rangle \text{ } &=&\left( 
\begin{array}{c}
{H}\left| \psi _{0}\right\rangle -\lambda _{1}({H}+\lambda _{2})\left|
p_{0}\right\rangle -\lambda _{2}\left| p_{0}\right\rangle \\ 
{H}\left| \psi _{1}\right\rangle -\lambda _{1}({H}+\lambda _{2})\left|
p_{1}\right\rangle -\lambda _{2}\left| p_{1}\right\rangle
\end{array}
\right) \ ,  \label{12} \\
\lambda _{1} &=&\left\langle Q|\Psi \right\rangle =\left\langle p_{0}|\psi
_{1}\right\rangle +\left\langle p_{1}|\psi _{2}\right\rangle ,  \nonumber \\
\lambda _{2} &=&\left\langle Q|{\bf V}_{{\rm S}}|\Psi \right\rangle
=\left\langle p_{0}|{V}_{{\rm S}}|\psi _{0}\right\rangle +\left\langle p_{1}|%
{V}_{{\rm S}}|\psi _{1}\right\rangle .  \nonumber
\end{eqnarray}
If the state $\left| \Psi \right\rangle $ belongs to the physical subspace
then, according to (\ref{8}), $\lambda _{1}=0$. However, the action of ${\bf %
PHP}$ even on the physical states is not reduced to that of ${\bf H}$
because the amplitude $\lambda _{2}$ is generally not zero. Thus, after the
projection (\ref{10}) the original Hamiltonian (\ref{6}) acquires an
additional term

\begin{equation}
{\bf PHP}=\left( {\bf H}+{\bf W}\right) P  \label{13}
\end{equation}

\begin{equation}
{\bf W}=-\left| Q\right\rangle \left\langle Q\right| \text{{\ }}{V}_{{\rm S}%
}.  \label{14}
\end{equation}

The operator ${\bf W}$ provides the channel mixing even if the initial state
is in the physical subspace. The correlation between the two components of $%
\left| \Psi \right\rangle $ is a consequence of the unbalanced action of the
operator $V_{{\rm S}}$ on the components of $\left| \Psi \right\rangle $. It
vanishes in the limit of the free negative ion.

\subsection{Interaction potentials}

The potential $V_{{\rm at}}$ in the Hamiltonian (\ref{6}) represents the
interaction between the active electron and the fluorine neutral core. It is
taken as a local model potential which depends on the distance $r$ between
the electron and the atom center. The potential also includes a long range
polarization interaction. Its explicit form has been adjusted to reproduce
the binding energy of the ion $F^{-}$ as well as the mean radius of the $p$%
-orbital. The potential reads 
\begin{eqnarray}
V_{{\rm at}} &=&-U_{0}+gr^{2}-ae^{-r^{2}}\ ,\ \ \ \ \ r\leq 1\ ;  \label{15}
\\
V_{{\rm at}} &=&-\frac{\alpha }{2r^{4}}-ae^{-r^{2}}\ ,\ \ \ \ \ \ \ \ \ \
r>1\ ,  \nonumber
\end{eqnarray}
where $U_{0}=5.64$, $g=3.76$, $a=1.2558$ and the atomic polarizability of a
fluorine atom $\alpha =3.76$ \cite{35}. All constants are given in the
atomic units.

The potential $V_{{\rm S}}$ in the Hamiltonian (\ref{6}) describes the
interaction of the active electron with the $Cu(111)$ and $Ag(111)$
surfaces. It has been proposed by Chulkov et al on the basis of their {\it %
ab initio} studies \cite{36}. This local potential depends only on the
electron coordinate $z$ along the surface normal. Its explicit form can be
found in Ref. \cite{36}. Qualitatively, it is an image charge potential in
vacuum which joins smoothly an oscillating potential with the period being
that of the $(111)$ planes inside the metal bulk. When describing an
electron motion in the direction perpendicular to the $Cu(111)$ or $Ag(111)$
surface, this potential represents rather well important features of the
surface such as the projected band gap (between -5.83 and -0.69 eV with
respect to vacuum for $Cu$ and between -4.96 and -0.66 eV for $Ag$), the
surface state (5.33 eV and 4.625 eV below vacuum for $Cu(111)$ and 
$Ag(111)$, 
respectively), and the image state energy positions (0.82 eV and 0.77 eV
below vacuum, respectively, for $Cu(111)$ and $Ag(111)$). Since the RCT
process mainly favors transitions around the surface normal, this potential
is expected to account for the effect of the pecularities of the $Cu(111)$
and $Ag(111)$ surfaces.

In order to illustrate the effects of the projected band gap of the $Cu(111)$
and $Ag(111)$ surfaces, we also present results obtained for a free-electron
description of the metal surface. In this case, the local electron-surface
interaction potential corresponds to the $Al(111)$ surface and is taken from
the work of Jennings et al \cite{37}.

\subsection{Wave packet propagation}

In the wave packet propagation (WPP) approach \cite{8,23}, one studies the
time evolution of an electron wave function $\left\langle \vec{r}\,|\Psi
(t)\right\rangle $ generated by the system Hamiltonian. Here we consider
both the static and dynamic problems. In the former, the distance between
the ion and the metal surface is fixed, while in the latter the ion collides
with the surface. In both cases, the initial state is the free ion state (%
\ref{5}).

The time evolution can be regarded as a sequence of infinitesimal time steps
generated by the evolution operator (\ref{11}) 
\begin{equation}
\left| \Psi (t+\Delta t)\right\rangle ={\bf U}(t,t+\Delta t)\left| \Psi
(t)\right\rangle .  \label{16}
\end{equation}
Since the initial state $\left| \Psi (t=0)\right\rangle =\left| \Psi
_{F^{-}}\right\rangle $ is orthogonal to the vector $\left| Q\right\rangle $%
, i.e., it is in the physical subspace, we can omit the projection operators
to the left and right of the exponential in (\ref{11}). So in (\ref{11}) we
take 
\begin{equation}
{\bf U}(t,t+\Delta t)=e^{-i\Delta {\bf PHP}}=e^{-i\Delta t({\bf H}_{{\rm at}%
}-E_{{\rm a}}\left| Q\right\rangle \left\langle Q\right| +{\bf PV_{{\rm S}}P}%
)}\ ,  \label{17}
\end{equation}
where the property ${\bf H}_{{\rm at}}\left| Q\right\rangle =E_{{\rm a}%
}\left| Q\right\rangle $ of the vector $\left| Q\right\rangle $ has been
used; $E_{{\rm a}}$ is the eigenvalue of $H_{{\rm at}}$ corresponding to the
eigenfunctions $p(r)Y_{l,m}(\theta ,\varphi )$ with $l=1$ and $m=0,\pm 1$.
For an infinitesimal time step $\Delta t$, the action of the evolution
operator (\ref{17}) can be evaluated by means of the split operator
approximation \cite{38,39} 
\begin{equation}
{\bf U}(t,t+\Delta t)=e^{-i\frac{\Delta t}{2}{\bf PV_{{\rm S}}P}}e^{-i\Delta
t({\bf H}_{{\rm at}}-E_{{\rm a}}\left| Q\right\rangle \left\langle Q\right|
)}e^{-i\frac{\Delta t}{2}{\bf PV_{{\rm S}}P}}+O(\Delta t^{3})\ .  \label{18}
\end{equation}
Making use of the commutation relation of ${\bf H}_{{\rm a}}$ and $\left|
Q\right\rangle \left\langle Q\right| $, this representation can be further
simplified 
\begin{equation}
{\bf U}(t,t+\Delta t)=e^{-i\frac{\Delta t}{2}{\bf PV_{{\rm S}}P}}\left\{
\left( e^{i\Delta tE_{{\rm a}}}-1\right) \left| Q\right\rangle \left\langle
Q\right| +{\bf I}\right\} e^{-i\Delta t{\bf H}_{{\rm at}}}e^{-i\frac{\Delta t%
}{2}{\bf PV_{{\rm S}}P}}+O(\Delta t^{3})\ .  \label{19}
\end{equation}
The action of the exponential involving $V_{{\rm S}}$ is evaluated via a
Taylor expansion in which four terms are typically kept for the time step $%
\Delta t=0.025$ atomic units.

The action of the kinetic energy operator is computed in the cylindrical
coordinates which are well suited to the symmetry of the problem 
\begin{equation}
T=-\frac{1}{2}\frac{\partial ^{2}}{\partial z^{2}}-\frac{1}{2\rho }\frac{%
\partial }{\partial \rho }\,\rho \,\frac{\partial }{\partial \rho }+\frac{%
m^{2}}{2\rho ^{2}}\equiv T_{z}+T_{\rho }\ ,  \label{20}
\end{equation}
where $T_{z}$ contains only the $z$-derivative. The exponential of ${\bf H}_{%
{\rm at}}$ in (\ref{19}) is then transformed as 
\begin{equation}
e^{-i\Delta t{\bf H}_{{\rm at}}}=e^{-i\Delta t{H}_{{\rm at}}}\,{\bf I}=e^{-i%
\frac{\Delta t}{2}(T_{z}+V_{{\rm at}})}e^{-i\Delta tT_{\rho }}e^{-i\frac{%
\Delta t}{2}(T_{z}+V_{{\rm at}})}\,{\bf I}+O(\Delta t^{3})\ .  \label{21}
\end{equation}
Recall that the operator ${\bf H}_{{\rm at}}$ is diagonal in the isotopic
two-dimensional space. So, the operator (\ref{21}) acts on both components
of $\left| \Psi (t)\right\rangle $ in the same way. Finally, all the
exponentials in (\ref{21}) are approximated by means of the Cayley
representation \cite{40} 
\begin{equation}
e^{-i\Delta tA}=\frac{1-i\frac{\Delta t}{2}\,A}{1+i\frac{\Delta t}{2}\,A}%
+O(\Delta t^{3})\ .  \label{22}
\end{equation}
In order to accurately reproduce the wave packet variation close to the atom
center, we use a mapping procedure \cite{41,42} defined by

\begin{eqnarray}
z &=&f(\xi )=0.05\xi +\frac{0.95\xi ^{3}}{400+\xi ^{2}}\ ,  \label{23} \\
\rho &=&f(\eta )=0.05\eta +\frac{0.95\eta ^{3}}{400+\eta ^{2}}\ ,  \nonumber
\\
&&\left\langle z,\rho |\Psi (t)\right\rangle =\frac{1}{\sqrt{\rho }}\tilde{%
\Psi}(t,z,\rho )\text{ }.  \nonumber
\end{eqnarray}

The wave packet $\tilde{\Psi}(t,z,\rho )$ is evaluated on a 2D mesh of
points $(\xi _{k},\eta _{j})$ of the size 1200$\times $800 with the step
size $\Delta =0.2$ atomic units for both the coordinates. At the grid
boundary, an absorbing potential is introduced \cite{43,44} in order to
avoid the wave packet reflection.

The kinetic energy operator (\ref{20}) has to be written in the auxiliary
variables $\xi $ and $\eta $ and then discretized. After the change of
variables (\ref{23}) the operator $T_{\rho }$ assumes the form 
\begin{equation}
T_{\rho }=-\frac{1}{2}\frac{1}{J\sqrt{f}}\frac{\partial }{\partial \eta }%
\frac{f}{J}\frac{\partial }{\partial \eta }\frac{1}{\sqrt{f}}+\frac{1}{2}%
\frac{m^{2}}{f^{2}}\ ,  \label{24}
\end{equation}
where $J(\eta )=f^{\prime }(\eta )$ is the Jacobian. The grid in the $\eta $%
-coordinate is set as $\eta _{j}=\Delta /2+\Delta (j-1)$. For every value $%
\xi $ we have $\tilde{\Psi}_{j}=\Psi (\xi ,\eta _{j})$, and the action of (%
\ref{24}) is defined by the following midpoint procedure 
\begin{equation}
\left( T_{\rho }\tilde{\Psi}\right) _{j}=-\frac{1}{2\Delta ^{2}}\frac{1}{%
J_{j}\sqrt{f_{j}}}\left[ \frac{f_{j+1/2}}{J_{j+1/2}}\left( \frac{\tilde{\Psi}%
_{j+1}}{\sqrt{f_{j+1}}}-\frac{\tilde{\Psi}_{j}}{\sqrt{f_{j}}}\right) -\frac{%
f_{j-1/2}}{J_{j-1/2}}\left( \frac{\tilde{\Psi}_{j}}{\sqrt{f_{j}}}-\frac{%
\tilde{\Psi}_{j-1}}{\sqrt{f_{j-1}}}\right) \right] +\frac{1}{2}\frac{m^{2}}{%
f_{j}^{2}}\ .  \label{25}
\end{equation}
Here the subscript $j\pm 1/2$ means that the corresponding function is taken
at the midpoint $\eta _{j}\pm \Delta /2$. A similar expression can be
obtained for the action of $T_{z}$ on the grid $\xi _{k}=\xi _{0}+\Delta
(k-1)$.

In the first series of calculations, we study the static problem when the
ion $F^{-}$ is at a fixed distance $Z$ from the metal surface. The survival
amplitude of the ion (the auto-correlation of the wave function) is given by 
\begin{equation}
A(t)=\left\langle \Psi (t=0)|\Psi (t)\right\rangle .  \label{26}
\end{equation}
{}From the Laplace transform of the function $A(t)$, one can obtain the
density of states (DOS) projected on the free ion wave function. The
structure of the DOS yields the energy level and its width for the negative
ion state interaction with the surface. In what follows, this width is
referred to as the ``static width'' to emphasize that it is extracted from
static calculations. It gives the electron transfer rate between the
negative ion and the metal surface in the static problem.

In the second series of calculations, we study the evolution of the electron
wave packet when a negative fluorine ion collides with the surface. The ion
is assumed to approach the surface along a straight line perpendicular to
the surface at a constant velocity $v$. Only the incoming part of the
collision is studied. The time dependence of the wave function is obtained
in the projectile reference frame, i.e., the time dependence of the
Hamiltonian occurs through the potential $V_{{\rm S}}$. For each collision
velocity, the ion survival probability, $P(t,v)=|A(t,v)|^{2}$, is computed.
To analyze the dynamics of the charge transfer, we define an effective width
of the negative ion state by 
\begin{equation}
G(Z,v)=-\frac{\partial \log [P(t,v)]}{\partial t}\ ,  \label{27}
\end{equation}
where $Z=Z_{0}-vt$ with $Z_{0}$ being an initial distance of the ion from
the metal surface. It corresponds to an effective decay rate of the ion when
it approaches the surface with a velocity $v$. Comparing $G(Z,v)$ to the
level width obtained in the static calculations allows us to see to what
extent the dynamical evolution can be described by the static width of the
ion level.

\newpage
\section{Results and discussion}

\subsection{${\bf F^-}$ ions interacting with a surface Al(111)}

The interaction of an $F^{-}$ ion with an $Al(111)$ surface, where the
latter is regarded as a free-electron metal surface, has already been
studied by the CAM method associated with the effective range treatment of
the negative ion \cite{12,13}. It lead to a successful description of the
negative ion formation in a grazing angle scattering \cite{12}. Similarly,
for large angle scattering from $Ag(110)$, and polycrystalline $Ag$ and $Al$
surfaces, a quantitative agreement with experiment results \cite{14,45} has
been obtained \cite{13}.

In Figures 1 and 2 we compare the results obtained in the static case by two
different methods: The CAM method with effective range treatment of the
negative ion and the present WPP results obtained with the projection
formalism. In both cases, the $Al(111)$ surface is described as a
free-electron metal surface using the potential proposed in \cite{37}.
Figures 1 and 2 present, respectively, the energy position and the width 
of the $F^{-}$
ion level interacting with the surface as a function of the ion-surface
distance $Z$ measured from the image plane. The characteristics of the ion
level as a function of $Z$ display the behavior common for atomic species in
front of a free-electron surface: The energy of the negative ion state
decreases as the ion is placed closer to the surface, which can be
anticipated because of the image charge attraction; The level width
increases roughly exponentially as $Z$ decreases. The results obtained by
two different methods are extremely close to each other. This gives
confidence in the equivalence of the two descriptions of the $F^{-}$ ion.
The results for a free-electron $Al(111)$ surface are used below as a ``free
electron'' reference to which compare the $Cu(111)$ and $Ag(111)$ results.
It appears that the free electron results are almost identical for the three
metals, except at very small distances from the surface.

\subsection{${\bf F}^-$ ions interacting with a surface Cu(111). A static
case}

Figures 3 and 4 present 
the $F^{-}$ ion level characteristics (energy (Fig.3) and width (Fig.4))
as a function of the ion-$Cu(111)$ surface distance. The negative ion level
energy exhibits an avoided crossing around $4a_{0}$ from the surface, which
is quite different from the smooth behaviour seen in Figures 1 and 2 for the
free-electron metal. This is a direct consequence of the peculiarities of
the $Cu(111)$ surface and a similar situation has already been observed in
the case of H- interacting with the same surface\cite{23}.

A schematic picture of the electronic structure of the model $Cu(111)$
surface is shown in Fig. 5. The energy of electronic levels is plotted as a
function of the electron momentum, $k_{\Vert }$, parallel to the surface.
For $k_{\Vert }$ equal to zero, the projected band gap lies within the
energy range from -5.83 to -0.69 eV (with respect to vacuum). Inside the
gap, there is a surface state at -5.33 eV. In the present model of a $Cu$
surface, the dispersion curves of all the metal electronic states as
functions of $k_{\Vert }$ are parabolic with a free-electron mass.

The resonant charge transfer process corresponds to transitions between the
ion level and metal states of the same energy. At large distances, the $%
F^{-} $ ion level is degenerate with the band gap. Therefore it can only
decay to metal states with a finite $k_{\Vert }$ i.e. either to the 2D
surface state band, or into 3D propagating states. As $Z$ decreases, the
energy of the negative ion state decreases and it comes close to the bottom
of the 2D surface state continuum. The ion can decay by ejecting an electron
with the angular momentum $m=0,\pm 1$ (the quantization axis is normal to
the surface). As explained in Ref. \cite{23}, a resonance cannot cross the
bottom of a 2D continuum in the symmetrical case $m=0$ and there always
exists a bound state below the bottom of the continuum. This state has an
avoided crossing with the state which becomes the free ion state as $Z$
tends to infinity. The $F^{-}$ ion character is then found to be associated
with two different states depending on the Z range. It is transferred from
the upper to the lower state when going through the crossing region
(decreasing $Z$). Far from the avoided crossing region, the ion energy level
is rather close to that found in the free electron case.

As for the width, at large $Z$, its absolute value for $F^{-}$ in front of
the $Cu(111)$ surface is larger than that in the case of a free-electron
surface. This result might appear surprising since the projected band gap
prohibits the electron transfer from the projectile to the metal along the
surface normal ($k_{\Vert }=0$) and blocks the RCT into the 3D bulk
continuum. Indeed, as can be seen in Fig. 5, there are no electronic states
of the metal with small $k_{\Vert }$ which are in resonance with the
negative ion state. The potential barrier separating the ion and the surface
attains its least value in the direction normal to the surface. Therefore,
the surface normal is the preferred direction of the resonant electron
transfer. One would then expect that the effect of the projected band gap
would be to stabilize the negative ion level as compared with the case of a
free electron metal. This has indeed been found for $H^{-}$ interacting with 
$Cu(111)$ where the width of the $H^{-}$ state was much reduced as compared
to the $H^{-}$/$Al(111)$ - case \cite{23}.

In contrast, we have observed an increase of the fluorine negative ion decay
rate as compared to the free-electron metal case. The reason is twofold.
First, the 2D surface state continuum contributes to the decay of $F^{-}$.
Second, a fluorine has a much larger electron affinity than a hydrogen.
Thanks to a better overlap of the wave functions, the 2D surface state
continuum, when energetically allowed, is a dominating decay channel for an
ion state lying within the band gap \cite{23,28,46}. Moreover, when the
binding energy of a negative ion is close to that of the surface state, a
coupling of the ion level with the 2D surface state continuum is more
efficient than its coupling with the 3D continuum of the free-electron metal
states (see a discussion in \cite{46}). The efficiency of the 2D surface
state continuum as a decay channel can also be deduced from the sharp
decrease of the level width when passing through the crossing region as $Z$
decreases, i.e., when this decay channel becomes closed.

For small values of $Z$: (i) the 2D surface state continuum does not
contribute to the decay of the negative ion, and (ii) the energy of the
negative ion state is very close to the bottom of the projected band gap for
small $k_{\Vert }$ so that the band structure effect for the decay into the
3D bulk continuum vanishes. Therefore, we find the width of the level with
ionic character very close to the free-electron results.

Finally, we can stress that the procedure of extracting the resonance
characteristics employed here is based on the autocorrelation function (\ref
{26}) using the free $F^{-}$ ion wave function as the initial state. It
converges well for the states of an ionic type. Convergence of the resonance
characteristics for other states is difficult to achieve. This is the reason
for showing only one state far from the crossing region (small or large $Z$%
). In the crossing region, the ionic character is shared between the lower
and upper states so that the characteristics (energy and width) of both of
them can be extracted. Since the convergence is easier to achieve for the
energy of the state, the interval of distances Z where both states are
presented is larger in Fig.3.

\subsection{${\bf F}^-$ ions interacting with an Ag(111) surface. Static and
dynamic studies}

The electronic structures of $Ag(111)$ and $Cu(111)$ look rather similar
(cf. Figures 5 and 6). However, characteristic features of the electronic
structures occur at different energies. The surface state in $Ag(111)$ is
located higher in energy than in $Cu(111)$. For this reason the avoided
crossing appears at a larger $Z$ where the ion-surface charge transfer
interaction is smaller. As a consequence the avoided crossing could not be
resolved in the energy dependence because it is localised in a too small
range of $Z$. Therefore, we have chosen to represent the results for the
energy and the width by a single continuous line (Figures 7 and 8, 
respectively). In fact, the energy of the
negative ion state is almost the same as for the free-electron metal
surface. The characteristic change of the level width when passing the
crossing region (decreasing $Z$) is however still perfectly visible. It
fully confirms the dominance of the 2D surface state channel in the $F^{-}$
ion decay at large Z.

At small $Z$, the $F^{-}$ ion level is embedded in the 3D propagating states
of $Ag$ and its characteristics are very close to those of the free-electron
case. Very close to the surface, the decrease of the level width as compared
to $Al(111)$ free-electron results is caused by the closeness of the $F^{-}$
level to the bottom of the $Ag(111)$ valence band. As can be seen in Fig. 3b
the bottom of the $Ag(111)$ valence band is located at $-9.7$ eV in our
model description 
of the $Ag(111)$ surface. For the model free-electron 
$Al(111)$ case it is located at $-15.9$ eV.

By studying the time dependent problem, one can find out whether the
peculiarities of $Ag(111)$ observed in the static case can still be visible
in a collision of $F^{-}$ with the $Ag(111)$ surface. We have computed the
effective level width $G$, defined by (\ref{27}), as a function of the
distance for an ion $F^{-}$ approaching the surface at different velocities.
The results are displayed in Fig. 9 together with the results of the static
case for $Ag(111)$ and the free-electron metal surface $Al(111)$. As the
collision velocity is increased, the effective width becomes closer to the
free-electron result. This feature is very similar to what we have found for
ions $H^{-}$ interacting with a $Cu(111)$ surface \cite{23}. The system
needs a finite time to react on the presence of the projected band gap. If
the collision is too fast, the electron wave packet does not have enough
time to ``explore'' the band structure of the metal, and the ion decay
remains identical to that on a free-electron metal surface.

In contrast, as the collision velocity is decreased, the effective width
comes nearer to the static $Ag(111)$ width. For the smallest velocity used
here, 0.0058 atomic units which correspond to a collision energy about 16
eV, the effective width is very close to the static one at large $Z$. When $%
Z $ is decreased, the effective width fails to perfectly reproduce the
change of the behavior associated with the crossing of the bottom of the
surface state continuum for all collision velocities considered here. We can
nevertheless see that this variation is better reproduced as the collision
velocity is decreased. In fact, the dynamical broadening introduced by the
change of the negative ion state energy with time \cite{47} makes it
impossible to reproduce a sharp variation of the static width as the ion
approaches the surface. This leads to rounded delayed structures displayed
in Fig.9. Moreover, the oscillations of the effective width at small $Z$ can
tentatively be attributed to the population transfer between the two
adiabatic states at the crossing region. From these results one can conclude
that in the studied collision energy range, 16 eV -- 5 keV, the charge
transfer rate is intermediate between the free-electron case and the static $%
Ag(111)$ case.

The formation of $F^{-}$ ions and $H^{-}$ ions by collision on a $Ag(111)$
surface has been studied experimentally by Guillemot and Esaulov\cite{29}.
When comparing with results obtained with free-electron like surfaces they
found that the $H^{-}$ data presents a strong band gap effect. In
particular, the survival probability of the negative ions leaving the
surface was much larger for $Ag(111)$. This was attributed to the blocking
of the resonant charge transfer in $H^{-}$/$Ag(111)$ system, in line with
theoretical results obtained for $H^{-}$/$Cu(111)$\cite{23} where the RCT
rates are reduced by orders of magnitude compared to the free-electron case.
At the same time, results obtained with $F^{-}$ ions were approximately
consistent with a description based on the charge transfer rate obtained in
the framework of a free-electron description of the metal surface. In the
energy range studied in their work, as shown above, the dynamical behaviour
of the charge transfer is intermediate between the free-electron and static $%
Ag(111)$ limits. This prohibits the use of the simple classical treatment of
the parallel velocity effect\cite{28,48} which was shown to be important for
a formation of $F^{-}$, even at rather low collision energy\cite{13}.
Consequently, we cannot quantitatively compare our results with theirs.
Nevertheless, we can notice that in our study the band gap effect is {\it %
reversed} and much smaller in the present system than in the $H^{-}/Cu(111)$
system; it is even partly suppressed by the ion motion. These findings
qualitatively agree with the experimental results\cite{29}.

\section{Conclusions}

We have reported on a study of the electron transfer in the ion-metal
systems $F^{-}/Cu(111)$ and $F^{-}/Ag(111)$. We have developed a new method
to describe the effect of six quasi-equivalent outer-shell electrons of $%
F^{-}$ on the resonant charge transfer process. The original six-electron
problem has been transformed into two one-electron problems in which the
dynamics are not independent but rather are linked via a constraint. The
projection formalism for quantum systems with constraints has been used to
obtain the quantum mechanical propagator for such a system. This modeling of
the ion $F^{-}$ is simple, efficient and can easily be implemented in the
wave packet propagation approach.

Both the $Cu(111)$ and $Ag(111)$ surfaces exhibit an electronic structure
with a projected band gap in which the ion energy level lies at large
ion-surface distances. This peculiarity of the electronic structure
influences the charge transfer interaction with the ion, leading to a few
remarkable features:

$\bullet $ The ion level presents an avoided crossing with the bottom of the
surface state continuum as predicted for a 2D continuum with $m=0$. The
avoided crossing is very clearly marked for the system $F^{-}/Cu(111)$.

$\bullet $ Because of the correlation between the six electrons of the $%
F^{-} $ ion the avoided crossing, which is a characteristic feature of the
symmetric case $m=0$, appears in the present system where electrons in both
states with $m=0$ and $|m|=1$ contribute to the charge transfer.

$\bullet $ When the negative ion level is low in the projected band gap, the
band gap does not cause a drastic drop of the charge transfer rate as
observed in other systems \cite{23,24}. This feature of the charge transfer
has been attributed to the following effects: ({\em i}) The band gap effect
is expected to decrease as the projectile level is lower in the band gap; (%
{\em ii}) The decay to the surface state is favored over the decay to 3D
bulk states because of a greater spatial overlap of the electron wave
function with the surface state; and ({\em iii}) The polarization of
negative ions does not enhance the band gap stabilization effect as it does
for neutral atoms (see a discussion in Ref. \cite{46}).

$\bullet $ At small ion-surface distances, when the negative ion state is
not in the band gap or inside the gap but close to its bottom, the energy
and the width of the negative ion level are practically identical to those
found in the free-electron surface case.

$\bullet $ Studies of the corresponding dynamical systems, when the ion $%
F^{-}$ approaches the surface, have shown that the above features survive,
although partially, over a large collision energy range.

\newpage
\begin{figure}
\vskip -3cm
\centerline{\psfig{figure=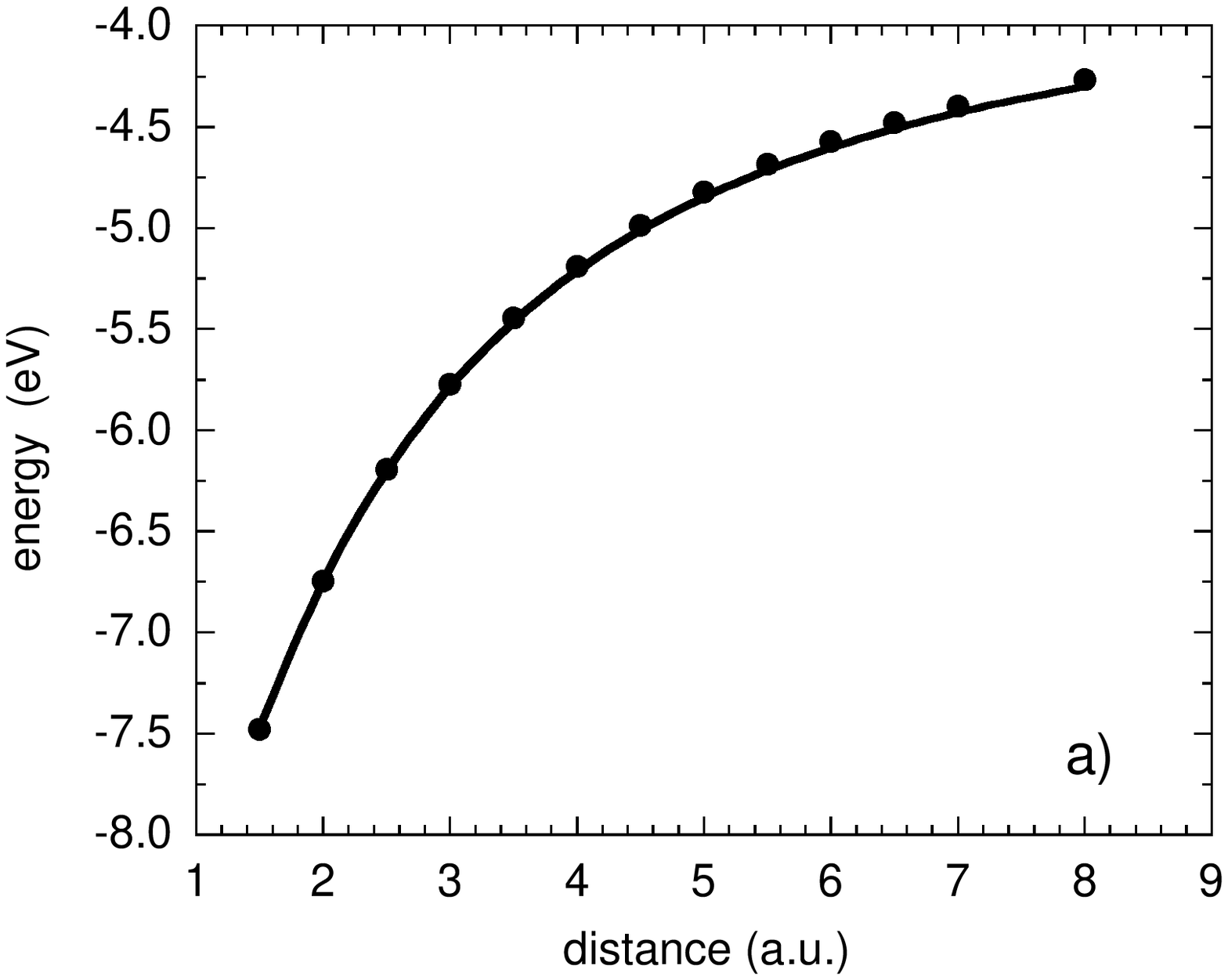,width=15cm,height=18cm}}
\vskip 1.5cm
\caption{Energy position of the $F^{-}$ ion
level in front of the $Al(111)$ free-electron-like surface, as functions of
the ion-surface distance, measured from the image plane (atomic units). The
solid line represents the results obtained with the present WPP approach.
The black dots indicate the results obtained with the CAM method.}
\end{figure}

\begin{figure}
\vskip -3cm
\centerline{\psfig{figure=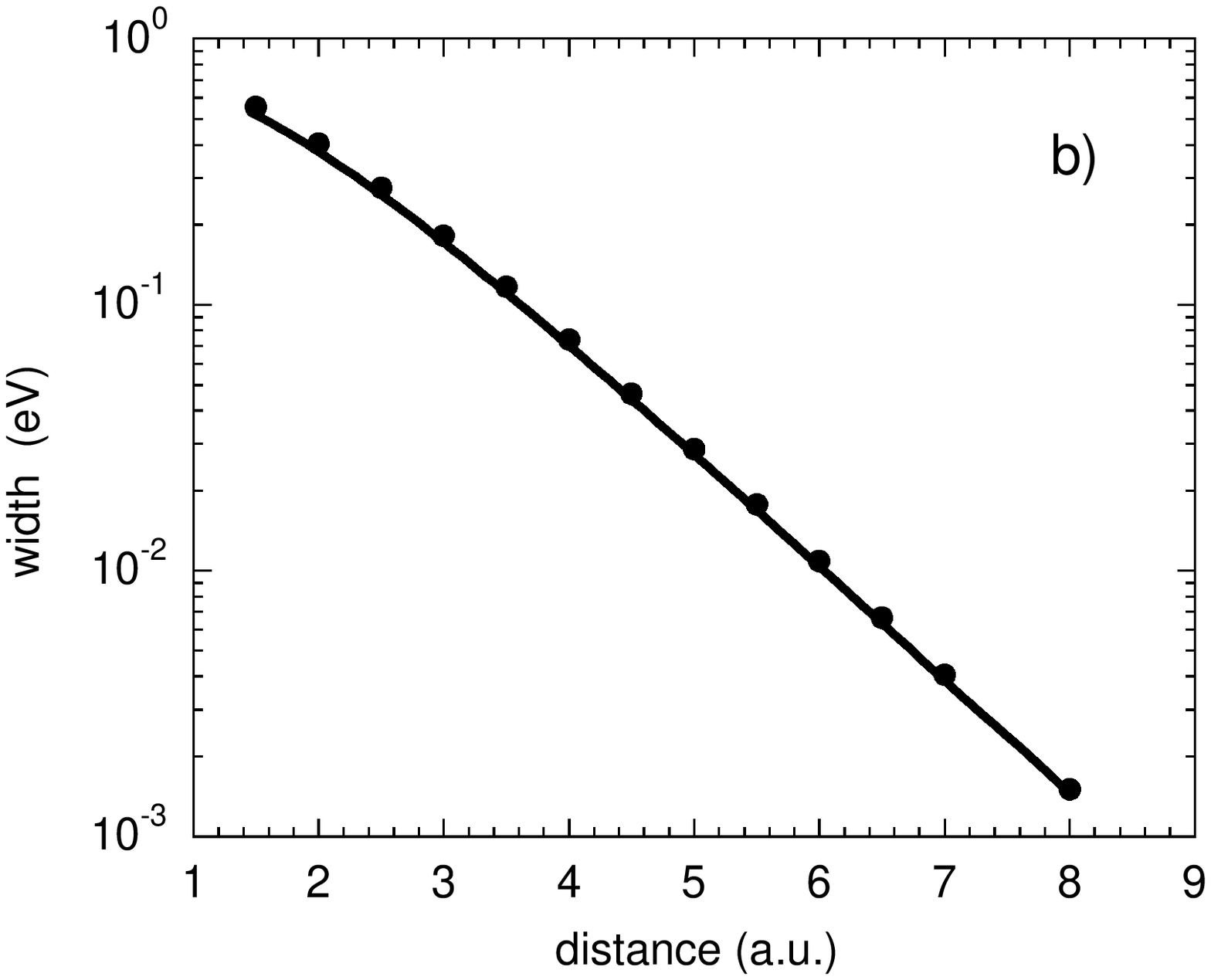,width=15cm,height=18cm}}
\vskip 1.5cm
\caption{Energy width for the same system as in Fig.1} 
\end{figure}

\begin{figure}
\vskip -3cm
\centerline{\psfig{figure=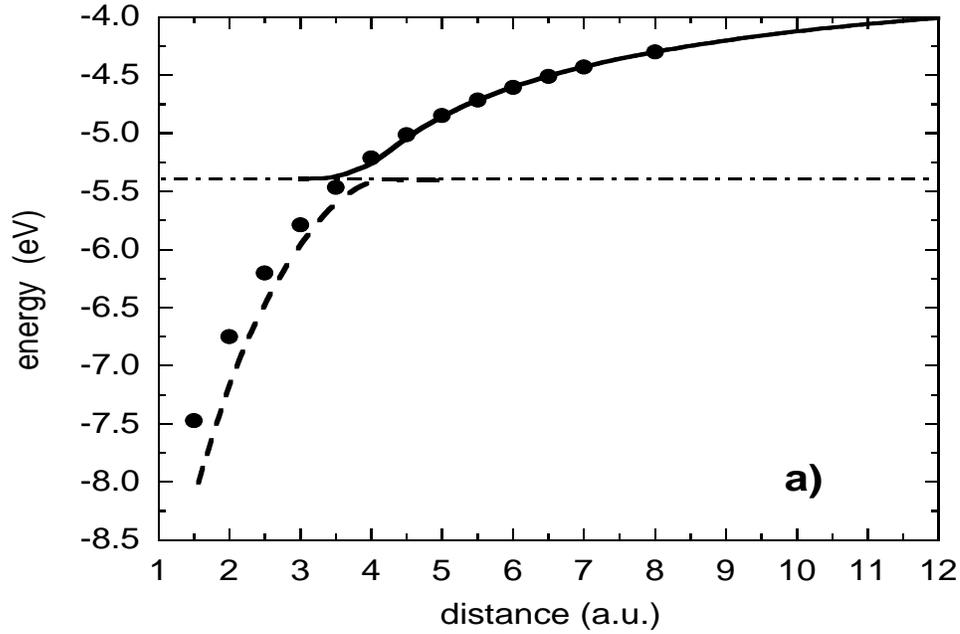,width=15cm,height=18cm}}
\vskip 1.5cm
\caption{ Energy position of the $F^{-}$ ion
level in front of the model $Cu(111)$ surface, as functions of the
ion-surface distance, measured from the image plane.(atomic units) The
energy reference is the vacuum level. Black dots: results for the
free-electron Al(111) surface. The horizontal dashed-dotted line indicates
the energy position of the bottom of the surface state continuum. Solid
line: results for the highest lying resonance. Dashed line: results for the
lowest lying resonance.} 
\end{figure}

\begin{figure}
\vskip -3cm
\centerline{\psfig{figure=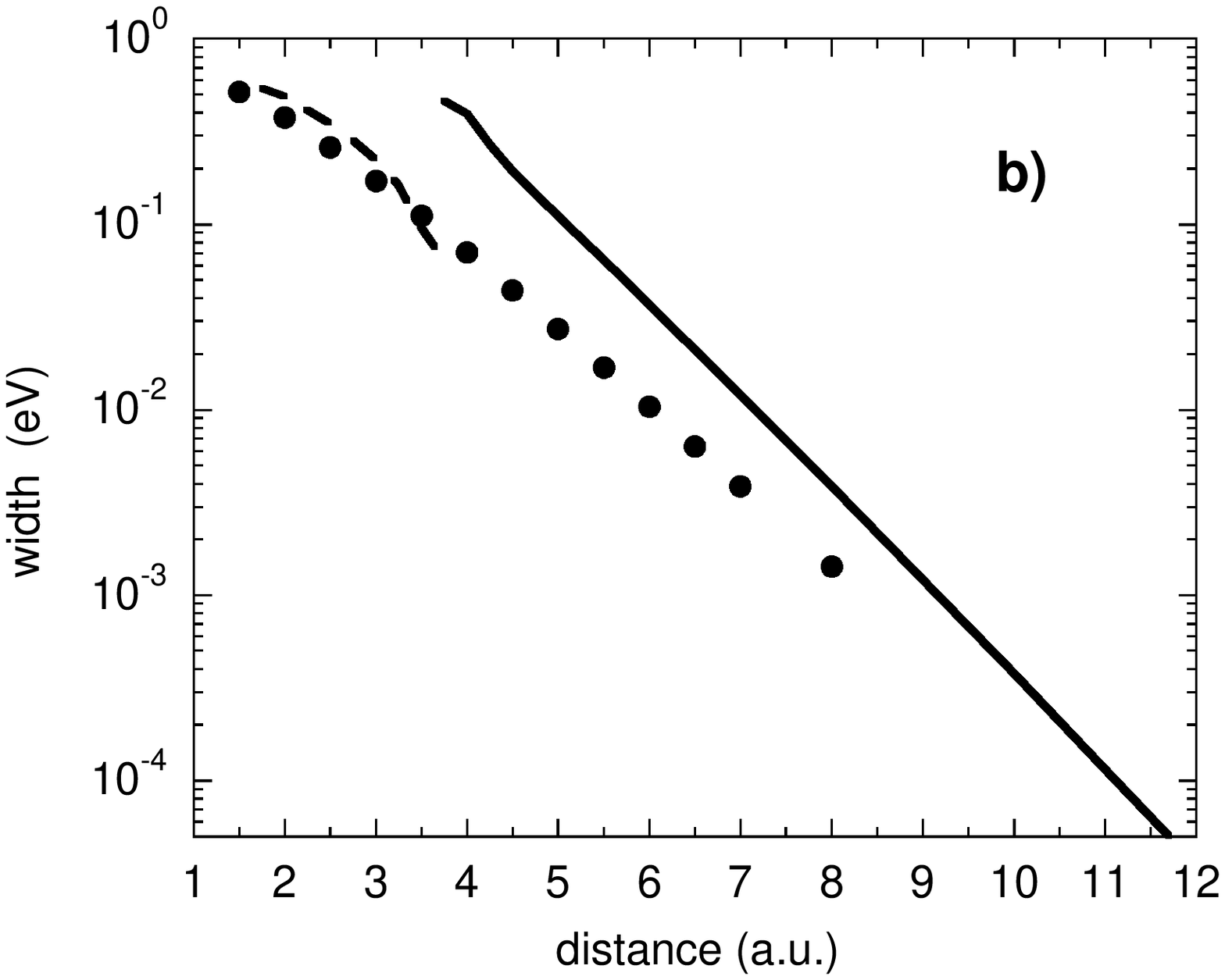,width=15cm,height=18cm}}
\vskip 1.5cm
\caption{ Energy width for the same system as in Fig.3.} 
\end{figure}

\begin{figure}
\vskip -3cm
\centerline{\psfig{figure=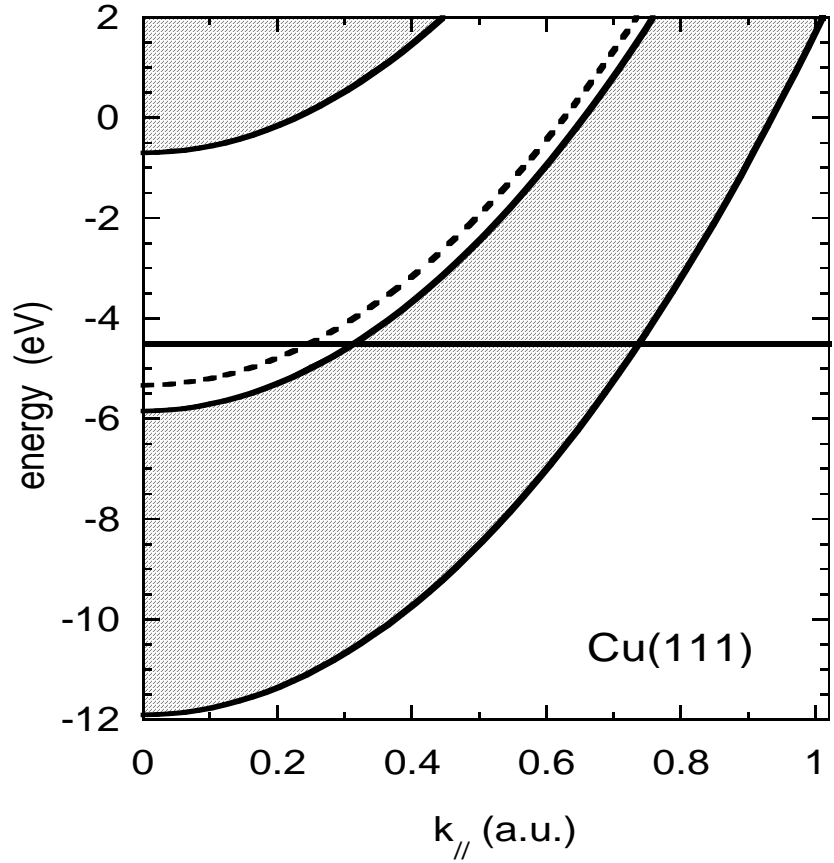,width=15cm,height=18cm}}
\vskip 1.5cm
\caption{ A schematic picture of the electronic structure of $%
Cu(111)$ (work function 4.9 eV) as a function of the electron momentum
parallel to the surface (atomic units). The energy reference is the vacuum
level. The shaded area represents the 3D valence band continuum. The dashed
line represents the 2D surface state continuum. The energy of $F^{-}$ level
at some distance from the surface is displayed as the horizontal solid line. } 
\end{figure}

\begin{figure}
\vskip -3cm
\centerline{\psfig{figure=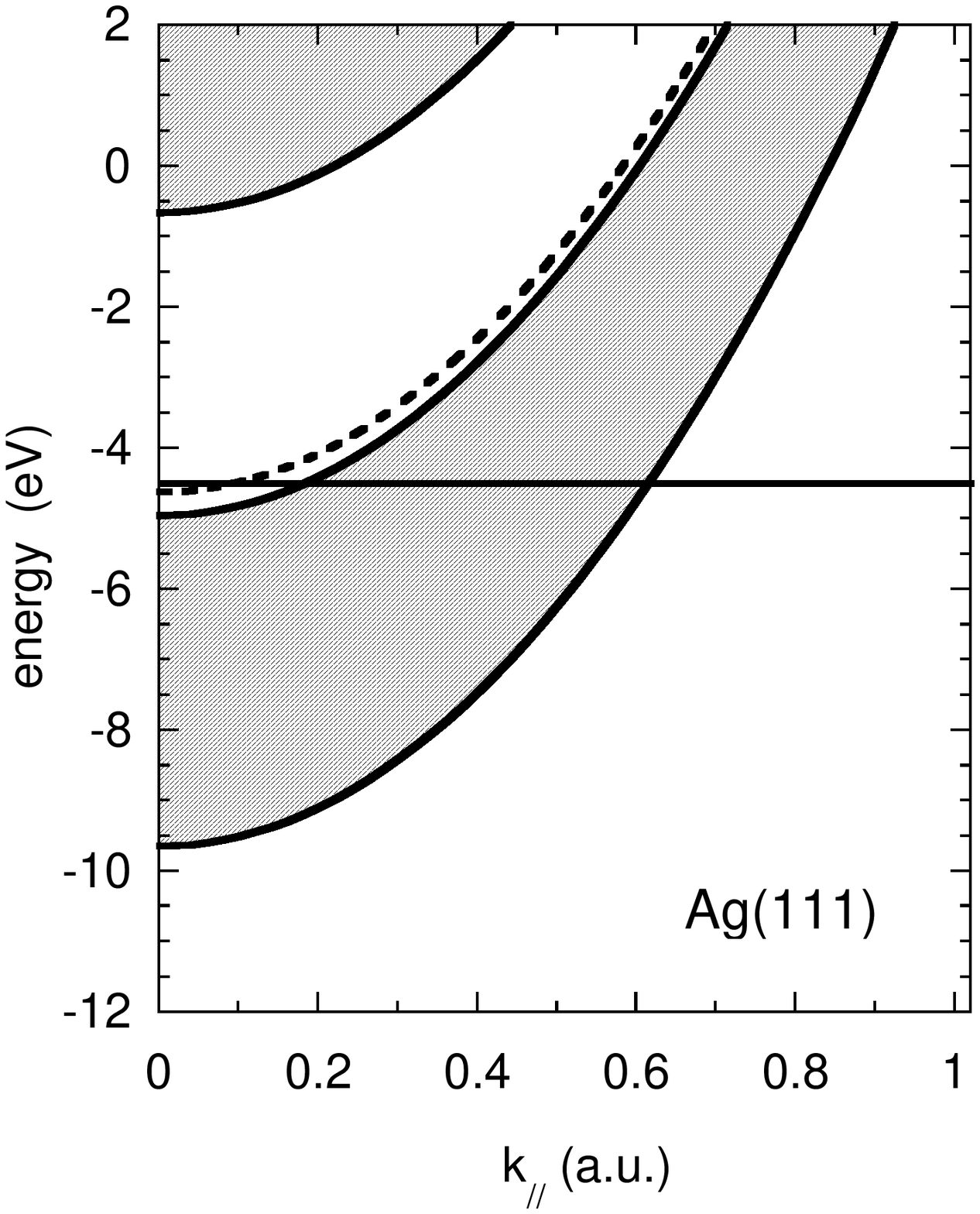,width=15cm,height=18cm}}
\vskip 1.5cm
\caption{ Same as Fig. 5 for the $Ag(111)$ surface (work
function 4.56 eV). } 
\end{figure}

\begin{figure}
\vskip -3cm
\centerline{\psfig{figure=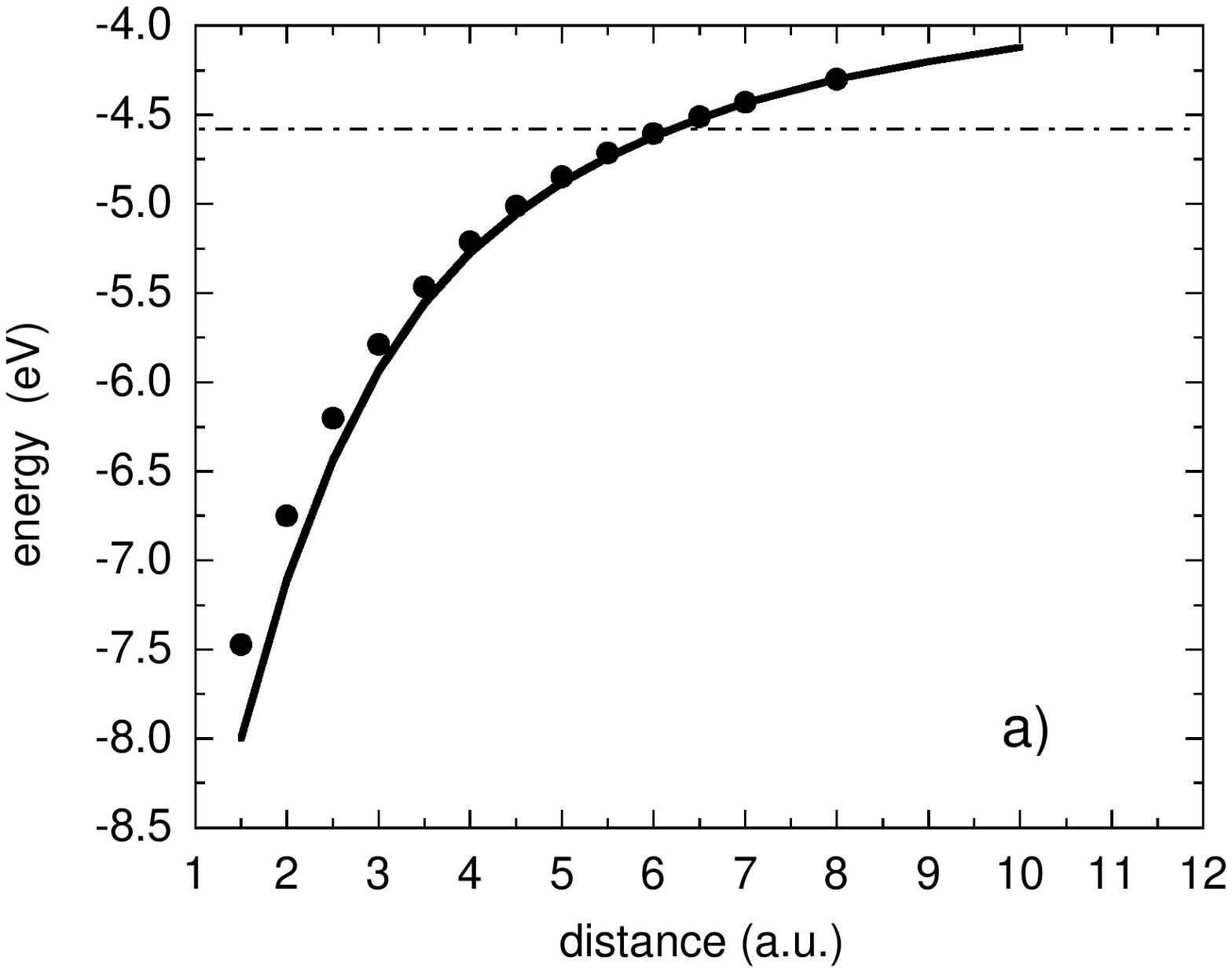,width=15cm,height=18cm}}
\vskip 1.5cm
\caption{ Energy position of the $F^{-}$ ion
level in front of the model $Ag(111)$ surface (solid lines), as functions of
the ion-surface distance, measured from the image plane (atomic units).
Black dots represent the results obtained for the free-electron $Al(111)$
surface. The horizontal dashed-dotted line indicates the energy position of
the bottom of the surface state continuum. } 
\end{figure}

\begin{figure}
\vskip -3cm
\centerline{\psfig{figure=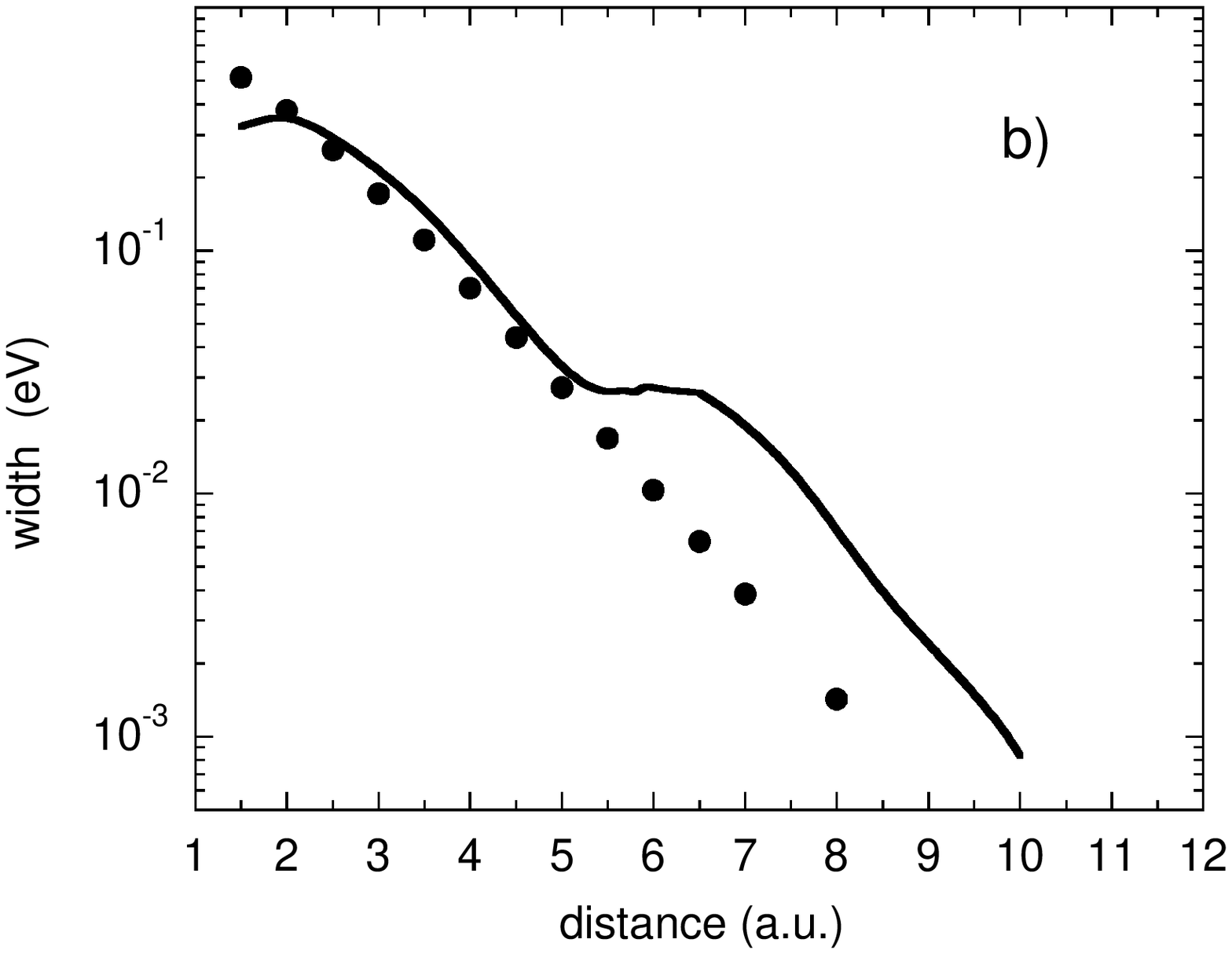,width=15cm,height=18cm}}
\vskip 1.5cm
\caption{Energy width for the same system as in Fig. 7.} 
\end{figure}

\begin{figure}
\vskip -3cm
\centerline{\psfig{figure=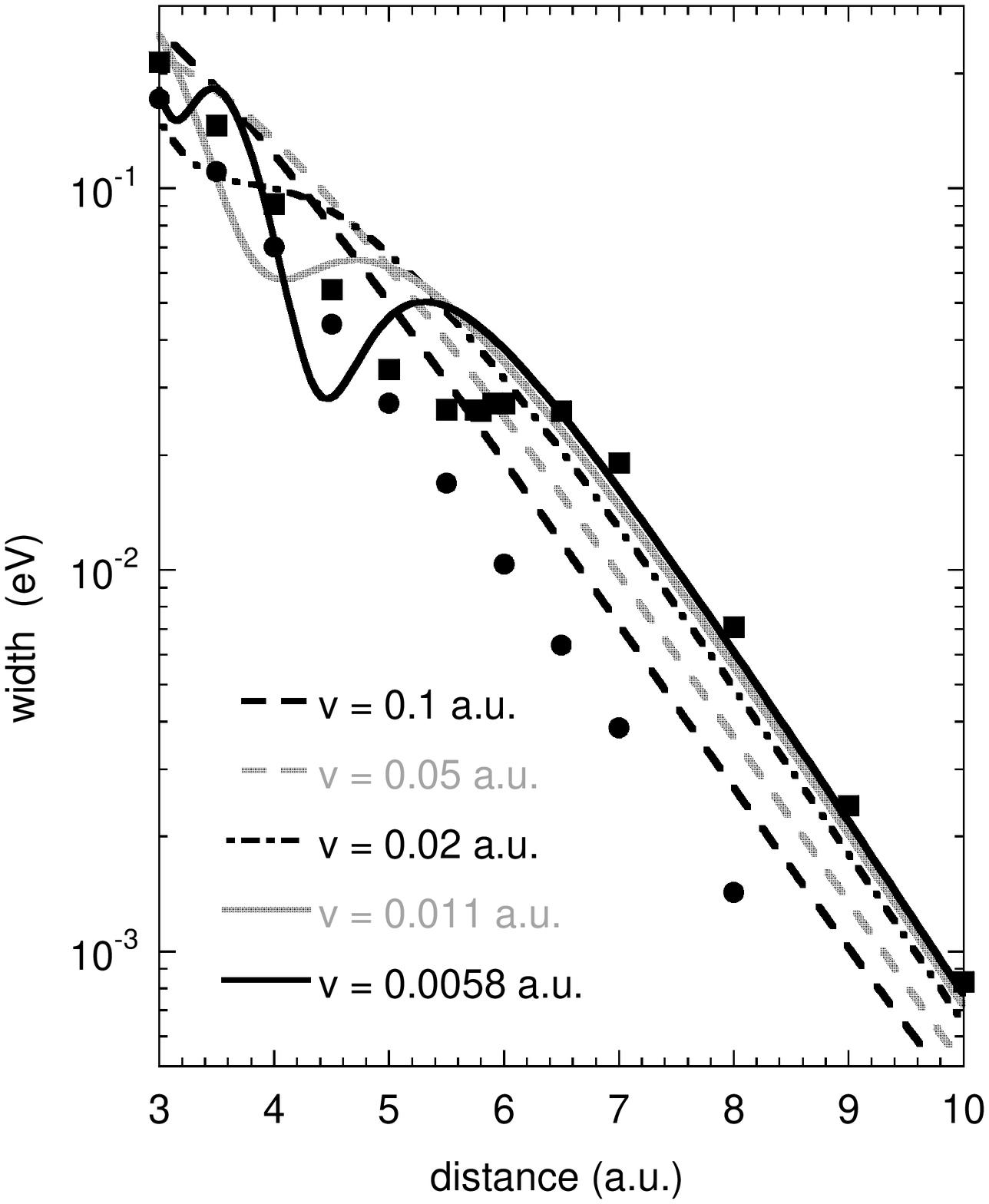,width=15cm,height=18cm}}
\vskip 1.5cm
\caption{The effective level width $G$ versus the ion-surface distance $Z$
for various collision velocities. The solid dots and solid squares stand
for, respectively, the free electron and $Ag(111)$ static widths. Continuous
lines represent the results obtained by the constrained wave packet
propagation method for the model surface $Ag(111)$ for various collision
velocities (see insert).} 
\end{figure}


\newpage

\begin{references}
\bibitem{1}  J.A. Misewich, T.F. Heinz and D.M. Newns, Phys. Rev. Lett. 68
(1992) 3737

\bibitem{2}  J.A. Prybyla, H.W.K. Tom and G.D. Aumiller, Phys. Rev. Lett. 68
(1992) 503

\bibitem{3}  L. Bartels et al., Phys. Rev. Lett. 80 (1998) 2004

\bibitem{4}  R. Hemmen and H. Conrad, Phys. Rev. Lett. 67 (1991) 1314

\bibitem{5}  P. Nordlander and J.C. Tully, Phys. Rev. B 42 (1990) 5564

\bibitem{6}  D. Teillet-Billy and J.P. Gauyacq, Surf. Sci. 239 (1990) 343

\bibitem{7}  F. Martin and M.F. Politis, Surf. Sci. 356 (1996) 247

\bibitem{8}  V.A. Ermoshin and A.K. Kazansky, Phys. Lett. A 218 (1996) 99

\bibitem{9}  J. Merino, N. Lorente, P. Pou and F. Flores, Phys. Rev. B 54
(1996) 10959

\bibitem{10}  S.A. Deutscher, X. Wang and J. Burgd\"orfer, Phys. Rev. A 55
(1997) 466

\bibitem{11}  P. K\"urpick, U. Thumm and U. Wille, Phys. Rev. A 56 (1997) 543

\bibitem{12}  A.G. Borisov, D. Teillet-Billy, J.P. Gauyacq, J.A.M.C. Silva,
A. Martens, C. Auth and H. Winter, Phys. Rev. B 59 (1999) 8218

\bibitem{13}  N. Lorente, A.G. Borisov, D. Teillet-Billy and J.P. Gauyacq,
Surf. Sci. 429 (1999) 46

\bibitem{14}  S. Ustaze, L. Guillemont, V.A. Esaulov, P. Nordlander and D.C.
Langreth, Surf. Sci. 415 (1998) L1027

\bibitem{15}  R. Zimny, H. Nienhaus and H. Winter, Radiation Effects and
Defects in Solids, 109 (1989) 9.

\bibitem{16}  J.P. Gauyacq, A.G. Borisov and H. Winter, Comments in Modern
Physics, Sect. D Atom. and Mol. Phys. 2 (2000) 29

\bibitem{17}  H. Shao, D.C. Langreth and P. Nordlander, in: Low energy
ion-surface interactions, Ed. by J.W. Rabalais (Wiley, New York, 1994) p. 118

\bibitem{18}  V.I. Arnold, V.V. Kozlov and A.I. Neishtadt, Mathematical
aspects of classical and celestial mechanics, in: Encyclodaedia of
Mathematical Science, Vol. III, Dynamical Systems (Springer-Verlag, Berlin,
1988)

\bibitem{19}  P.A.M. Dirac, Lectures on Quantum Mechanics (Yeshiva Univ.,
New York, 1964)

\bibitem{20}  L.D. Faddeev, Theor. Math. Phys. 1 (1970) 1.

\bibitem{21}  S.V. Shabanov, Phys. Reports 326 (2000) 1

\bibitem{22}  S.V. Shabanov, in: Proiceeding of ``Path Integrals'96'', Eds.
V.S. Yarunin and M.A. Smondyrev, (JINR, Dubna, 1996) p. 133; LANL electronic
archive, quant-ph/9608018; ``Path Integral in Holomorphic Representation
without Gauge Fixation'', JINR preprint E2-89-678 (Dubna, 1989)
(unpublished).

\bibitem{23}  A.G. Borisov, A.K. Kazansky and J.P. Gauyacq, Phys. Rev. Lett.
80 (1998) 1996; Phys. Rev. B59 (1999) 10935

\bibitem{24}  A.G. Borisov, A.K. Kazansky and J.P. Gauyacq, Surf. Sci. 430
(1999) 165

\bibitem{25}  M.C. Desjonqu\`{e}res and D. Spanjaard, Concepts in Surface
Science, Springer-Verlag Series in Surface Science, Vol. 40
(Springer-Verlag, Berlin, 1993)

\bibitem{26}  M. Bauer, S. Pawlik and M. Aeschlimann, Phys. Rev. B 55 (1997)
10040

\bibitem{27}  S. Ogawa, H. Nagano and H Petek, Phys. Rev. Lett. 82 (1999)
1931

\bibitem{28}  T. Hecht, H. Winter, A.G. Borisov, J.P. Gauyacq and A.K.
Kazansky, Phys. Rev. Lett. 84 (2000) 2517

\bibitem{29}  L. Guillemot and V.A. Esaulov, Phys. Rev. Lett. 82 (1999) 4552

\bibitem{30}  L. Guillemot, S. Lacombe and V.A. Esaulov, Nucl. Inst. Meth. B
164-165 (2000) 601

\bibitem{31}  D. Teillet-Billy, L. Malegat and J.P. Gauyacq, J. Phys. B 20
(1987) 3201

\bibitem{32}  S.V. Shabanov, J. Phys. A24 (1991) 1199; and Phase Space
Structure in Gauge Theories, JINR series of lectures, Vol. 54 (JINR, Dubna,
1989), P2-89-553

\bibitem{33}  J.R. Klauder, Ann. Phys. (NY) 254 (1997) 419; Nucl. Phys. B547
(1999) 397

\bibitem{34}  J.R. Klauder and S.V. Shabanov, Phys. Lett. B 398 (1997) 116;
Nucl. Phys. B511 (1998) 713; Phys.Lett. B435 (1998) 343

\bibitem{35}  Handbook of Chemistry and Physics, 5th Edition (CRC, Boca
Raton, Florida, 1996)

\bibitem{36}  E.V. Chulkov, V.M. Silkin and P.M. Echenique, Surf. Sci. 437
(1999) 330

\bibitem{37}  P.J. Jennings, P.O. Jones and M. Weinert, Phys. Rev. B 37
(1988) 6113

\bibitem{38}  M.D. Fleit and J.A. Fleck, J. Chem. Phys. 78 (1982) 301

\bibitem{39}  R. Kosloff, J. Phys. Chem. 92 (1988) 2087

\bibitem{40}  R.J. Taylor, Scattering Theory; the Quantum Theory of
non-Relativistic Collisions (R.E. Krieger Publishing Company, Malabar,
Florida, 1983)

\bibitem{41}  E. Fattal, R. Baer and R. Kosloff, Phys. Rev. E 53 (1996) 1217

\bibitem{42}  V. Kokoouline, O. Dulieu, R. Kosloff and F. Masnou-Seeuws, J.
Chem. Phys. 110 (1999) 9865

\bibitem{43}  R. Kosloff and D. Kosloff, J. Comp. Phys. 63 (1986) 363

\bibitem{44}  D. Neuhauser and M Baer, J. Chem. Phys. 91 (1989) 4651

\bibitem{45}  M. Maazouz, S. Ustaze, L. Guillemot and V.A. Esaulov, Surf.
Sci. 409 (1998) 189

\bibitem{46}  J.P. Gauyacq, A.G. Borisov, G. Raseev and A.K. Kazansky,
Faraday Discussions 117 (2000) 15

\bibitem{47}  J.J.C. Geerlings, J. Los, J.P. Gauyacq and N.M. Temme, Surf.
Sci. 172 (1986) 257

\bibitem{48}  J.N.M. van Wunnik, R. Brako, K. Makoshi and D.M. Newns, Surf.
Sci. 126 (1983) 618

\end{references}
\end{document}